\documentclass[twocolumn]{aastex631}
\usepackage{rotating}
\usepackage{booktabs}
\usepackage{amsmath,amsthm,amsfonts,amssymb,extarrows}
\usepackage{apjfonts}
\usepackage{float}
\usepackage{xcolor}
\usepackage{xspace}
\usepackage{longtable}

\definecolor{mBlue}{RGB}{51, 77, 167}

\hypersetup{linkcolor=red,citecolor=mBlue,filecolor=cyan,urlcolor=mBlue}

\received{\today}

\shorttitle{Fe K-shell DR satellites}
\shortauthors{Chintan~Shah~{et~al.}}

\begin{document}

\title{\Large{{Comprehensive Laboratory Benchmark of K-shell Dielectronic Satellites of\\ \ion{Fe}{25} --\ion{}{21} Ions}}}

\correspondingauthor{Chintan Shah}
\email{chintan.shah@mpi-hd.mpg.de}

\author[0000-0002-6484-3803]{Chintan~Shah}
\affiliation{NASA Goddard Space Flight Center, 8800 Greenbelt Rd, Greenbelt, MD 20771, USA}
\affiliation{Max-Planck-Institut f\"ur Kernphysik, Saupfercheckweg 1, 69117 Heidelberg, Germany}
\affiliation{Department of Physics and Astronomy, Johns Hopkins University, Baltimore, Maryland 21218, USA}

\author[0000-0002-5257-6728]{Pedro~Amaro}
\affiliation{Laboratory of Instrumentation, Biomedical Engineering and Radiation Physics (LIBPhys-UNL), Department of Physics, NOVA School of Science and Technology, NOVA University Lisbon, 2829-516 Caparica, Portugal}

\author[0000-0001-5777-1891]{Filipe~Grilo} 
\affiliation{Laboratory of Instrumentation, Biomedical Engineering and Radiation Physics (LIBPhys-UNL), Department of Physics, NOVA School of Science and Technology, NOVA University Lisbon, 2829-516 Caparica, Portugal}

\author[0000-0001-9136-8449]{Ming~Feng~Gu}
\affiliation{Space Science Laboratory, University of California, Berkeley, CA 94720, USA}

\author[0000-0001-9911-7038]{Liyi~Gu}
\affiliation{SRON Netherlands Institute for Space Research, Niels Bohrweg 4, 2333 CA Leiden, the Netherlands}
\affiliation{Leiden Observatory, Leiden University, P.O. Box 9513, 2300 RA Leiden, The Netherlands }

\author[0000-0002-5890-0971]{Jos\'e~Paulo~Santos}
\affiliation{Laboratory of Instrumentation, Biomedical Engineering and Radiation Physics (LIBPhys-UNL), Department of Physics, NOVA School of Science and Technology, NOVA University Lisbon, 2829-516 Caparica, Portugal}

\author[0000-0002-6374-1119]{F.~Scott~Porter}
\affiliation{NASA Goddard Space Flight Center, 8800 Greenbelt Rd, Greenbelt, MD 20771, USA}%

\author[0000-0002-5312-3747]{Thomas~Pfeifer}
\affiliation{Max-Planck-Institut f\"ur Kernphysik, Saupfercheckweg 1, 69117 Heidelberg, Germany}

\author[0000-0002-3331-7595]{Maurice~A.~Leutenegger}
\affiliation{NASA Goddard Space Flight Center, 8800 Greenbelt Rd, Greenbelt, MD 20771, USA}

\author[0000-0002-2937-8037]{Jos\'e~R.~{Crespo~L\'opez-Urrutia}}
\affiliation{Max-Planck-Institut f\"ur Kernphysik, Saupfercheckweg 1, 69117 Heidelberg, Germany}

\begin{abstract}
We report on comprehensive laboratory studies of the K-shell dielectronic recombination (DR) resonances of \ion{Fe}{25} -- \ion{}{21} ions that prominently contribute to the hard X-ray spectrum of hot astrophysical plasmas. By scanning a monoenergetic electron beam to resonantly excite trapped Fe ions in an electron beam ion trap, and achieving a high electron-ion collision energy resolution of $\sim$7~eV, we resolve their respective KL$n$ satellites up to $n^\prime = 11$. By normalization to known radiative recombination cross sections we also determine their excitation cross sections and that of the continuum with uncertainties below 15\%, and verify our results with an independent normalization based on previous measurements. Our experimental data excellently confirm the accuracy and suitability of distorted-wave calculations obtained with the Flexible Atomic Code (FAC) for modeling astrophysical and fusion plasmas.
\end{abstract}

\section{Introduction}\label{sec:intro}

Most of the baryonic matter in the Universe appears in hot ionized plasmas (10$^{5}$--10$^{7}$ K) that emit X-rays. They fill intergalactic space \citep{Reimers2002,Bykov2008,Nicastro2008,nicastro2018observations}, circumgalactic and interstellar media, and the interiors of stars. Thus, in the last few decades X-ray astronomy has become indispensable for exploring the high-energy universe and the extreme physical conditions prevalent in such astrophysical objects (see~\citet{fabian2023} and references therein). The most prominent and intensively studied features in X-ray spectra are the K-shell emission lines of Fe~\citep{asplund2021} and especially, the K$\alpha$ of helium-like~\ion{Fe}{25} at $\sim$6.7~keV. It is observed by missions such as ASCA, Chandra, XMM-Newton, Suzaku, Hitomi, {and XRISM}~\citep{ebisawa1996,murnion1996,sterling1997,peterson2001,tamura2001,kaastra2001,bianchi2005,nandra2006,miller2008,hitomi2016,hitomi2018,xrism2024cygx3,xrism2024ngc4151} in active galactic nuclei, supernova remnants, galaxy clusters, and solar flares. The K$\alpha$ lines from this Fe ion and other lower charge states serve as critical diagnostics for temperature, ionization state, and elemental abundances in these energetic sources~\citep{porquet2010,decaux1995,decaux1997,rudolph2013,steinbruegge2022}.

\begin{figure*}
    \centering
    \includegraphics[width=\linewidth]{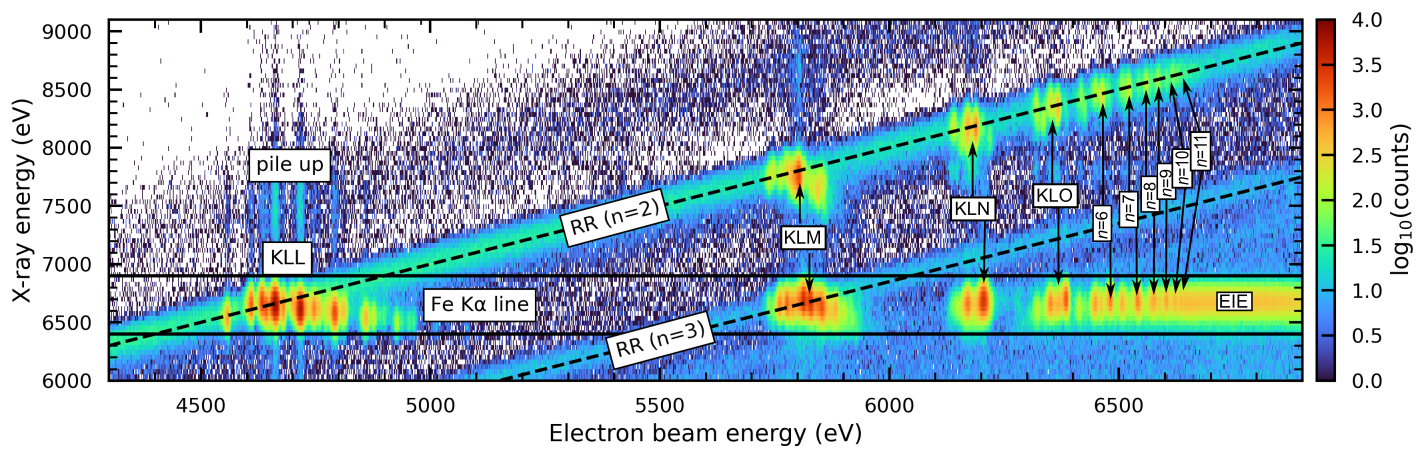}
    \caption{Intensity histogram of energy-resolved X-ray emission of Fe ions versus electron-beam energy. The solid lines contain the DR satellite channels of K$\alpha$ at $\sim$6.7 keV resolved up to $n=11$. Two diagonal dashed lines mark radiative recombination (RR) into $n=2$ and 3 shells. Above the RR $n=2$ line, artifacts due to their pile up with L-shell emissions below 2 keV are visible.}
    \label{fig:exp2d}
\end{figure*}

The formation of X-ray spectral lines in such environments is governed by a complex interplay of direct processes, such as electron-impact excitation (EIE), and indirect ones, such as dielectronic recombination, resonance excitation, and charge exchange~\citep{beiersdorfer1995,beiersdorfer2003}. In hot plasmas, dielectronic recombination (DR) dominates line formation with cross sections several orders of magnitude higher than those of EIE~\citep{knapp1989,knapp1993}, and strongly affects the charge-state balance of the plasma~\citep{dubau1980,beiersdorfer1992a}. 
{First described by~\citet{massey1942}, its importance was recognized by~\citep{burgess1964} for dynamics of hot solar corona plasmas.} 
DR is a two-step resonant mechanism in which a free electron is captured by a positive ion while simultaneously exciting a bound electron. Such doubly excited states decay by emitting distinctive satellite lines in dependence of the electron-energy distribution in resonance with them. Thus, DR satellites serve as sensitive probes of temperature and density, even for non-Maxwellian electron distributions~\citep{widmann1995,beiersdorfer1992a,beiersdorfer1993,kato1998,radtke2000}. 

Recent astrophysical observations have highlighted both the potential and the challenges of DR-based diagnostics. The high-resolution spectrum of the Perseus cluster obtained by the Hitomi X-ray Observatory in 2016~\citep{hitomi2016} revealed serious discrepancies between often used atomic databases and spectral fitting codes such as atomDB/APEC~\citep{foster2012}, SPEX~\citep{kaastra1996}, and CHIANTI~\citep{delzanna2015}. The observed K$\alpha$ complex of~\ion{Fe}{25} consists of four main transition lines ($w$, $x$, $y$, and $z$) excited by EIE with mostly unresolved contributions from DR satellites. Inconsistencies in their positions and cross sections across the aforementioned models led to 15-20\% variations in the derived plasma parameters~\citep{hitomiatomic2018}. Furthermore, high-$n\geq3$ DR satellites blend on the red side of the resonance line $w$, and can easily cause apparent line broadening and shifts of K$\alpha$. Without proper accounting, this introduces inaccuracies in the analysis of temperature, velocity and turbulence in astrophysical and laboratory fusion plasmas~\citep{beiersdorfer1992b}. Improving our understanding of DR processes is thus essential to fully exploit the capabilities of X-ray observatories equipped with high-resolution calorimeter-based missions, such as XRISM~\citep{xrism2018}, which was launched last year and has already started taking observations~\citep{xrism2024cygx3,xrism2024ngc4151}, and the upcoming Athena~\citep{pajot2018athena}. 

In the last decades, many experimental and theoretical studies of DR of highly charged ions (HCI) were reported. Intra-shell ({$\Delta n=0$}) low-energy DR rate coefficients, important for low-temperature photoionized plasmas, were measured at storage rings~\citep{savin2002a,savin2002b,schnell2003,savin2007,orban2010}, while inter-shell ({$\Delta n=1$}) DR strengths and rate coefficients, which are crucial for high-temperature collisionally ionized plasmas, were studied in electron beam ion traps (EBIT)~\citep{knapp1989,knapp1993,fuchs1998,smith2000,zhang2004,orourke2004,gonzalez2005,nakamura2008,yao2010,beilmann2009,ali2011,beilmann2011,hu2013,shah2015,shah2016,amaro2017,shah2018,shahharman2019}. DR measurements on highly charged Fe ions mostly focused on the resonance strengths of entire regions (KLL, KLM) of heliumlike~\ion{Fe}{25} and could not resolve neither level nor satellites beyond $n^\prime \geq 6$, and were not charge-state specific~\citep{beiersdorfer1992a,beiersdorfer1992b,beiersdorfer1993,watanabe2001}. Here, KLL denotes in analogy to the time-reversed process of Auger-Meitner decay the dielectronic capture process where a free electron recombines into the L-shell by resonantly exciting a bound electron from the K to the L shell. 

We fill these critical gaps in essential experimental data by targeted comprehensive measurements of K-shell DR resonances of~\ion{Fe}{25}\,--\ion{}{21} ions with an EBIT. Our outstanding electron-beam energy resolution of about 7~eV allows us to resolve KL$n$ satellites up to $n^\prime$ = 11. With these data, we determined DR resonance strengths with uncertainties below $\sim$15\%. Our results agree overall very well with atomic structure and distorted wave calculations performed with the Flexible Atomic Code (FAC)~\citep{gu2008} including EIE. Our accurate benchmarking of this code recommends its use in generating the atomic data sets needed for astrophysical and fusion plasma models.

\section{Measurements}\label{sec:exp}

{We employ FLASH-EBIT \citep{epp2007,epp2010} for the present experiment.} In this device, a nearly monoenergetic electron beam is produced by a barium oxide dispenser cathode, with currents ranging from tens to hundreds of milliamperes. The beam is compressed by a 6-T superconducting magnet in Helmholtz configuration to a diameter of less than 50~$\mu$m~\citep{shah2018}. Iron was continuously injected into the trap region in the form of iron pentacarbonyl (Fe(CO)$_5$) molecules through a two-stage differential pumping system. After dissociation of this molecule by the electron beam, the Fe atoms experience successive electron-impact ionization yielding HCI which are radially trapped in the negative space charge potential of the electron beam and axially confined by electrostatic potentials applied to the cylindrical drift-tube electrodes surrounding the trap region.

We swept the electron-beam energy over a range of KL$n$ resonances, whereby a free electron is captured in one of the $n$ shells upon exciting a bound electron from the K- to the L-shell. For this we vary the drift tube platform voltage, which biases all the drift tubes, from 2.1 to 5.1~kV in a triangular wave pattern at a slew rate of 50~V\,s$^{-1}$. This, in addition to the 2~kV negative bias applied to the cathode yields electrons of the required kinetic energies. {In order to enhance the electron-beam energy resolution, we reduce the nominal axial trapping potential to only 10~V while keeping the current at 100~mA to ensure enough production of Fe HCI. We optimize the axial trap depth, beam current, together with the Fe(CO)$_5$ injection pressure to force evaporative cooling of the trapped ions, resulting in a small characteristic ion gyroradius which samples only a fraction of the space charge distribution of the electron beam~\citep{penetrante1991a,beilmann2009,shah2016}.} This gives us a relative electron-energy resolution of $E/\Delta E\approx900$ at 6.5~keV, nearly ten times better than earlier Fe works~\citep{beiersdorfer1992a,beiersdorfer1992b,watanabe2001}. {This allows us to observe the DR satellites maximum up to $n\prime$ = 11 contributing to Fe K-shell X-ray lines (see Fig.~\ref{fig:exp2d})}. We dump the ion content of the trap for 3~s at the end of each 60~s sweep cycle to prevent a slow accumulation of Ba and W ions generated from atoms emanating from the hot cathode.

X-rays from the radiative decay of doubly excited KL$n$ DR resonances were recorded over 20 hours using a silicon drift detector (SDD) with an energy resolution of $\sim$120~eV FWHM at $\sim$6~keV while sweeping the electron beam energy. The resulting two-dimensional (2D) intensity histogram of X-ray energy versus electron beam energy is shown in Fig.~\ref{fig:exp2d}. The diagonal bands visible there correspond to radiative recombination (RR) into states with principal quantum numbers $n=2$ and $n=3$. Here, the X-ray energy is the sum of free electron energy and binding energy of the L- and M-shell, respectively. Bright spots between the solid lines encompassing X-rays of about 6.7 keV are due to the $n=2\rightarrow1$ radiative decay following the formation of doubly excited KL$n$ DR states, where $n=2$,\,3,\,4,\,etc. The clustered resonances between 4500 and 5000~eV beam energy arises from KLL DR resonances of \ion{Fe}{25}\,--\ion{}{21} ions. Resonances from lower charge states of Fe appear at lower photon energies since more bound electrons screen the nuclear potential for the $n=2$ electrons. At 5800~eV electron-beam energy, the KLM resonances appear, which can radiatively decay either by $n=2\rightarrow1$ or $n=3\rightarrow1$ thus emitting K$\alpha$ or K$\beta$ photons, respectively. 
{Analogous radiative decay pathways of higher KL$n$, where $n$ is up to 11, DR resonances appear at higher beam energies, and merge above a threshold around 6.5\,keV with the direct EIE of the $1s-np$ transitions.}

\section{Theoretical Calculations}\label{sec:th}
\begin{table*}
  \centering
  \footnotesize
  \caption{Present FAC predictions of KLL DR resonance and X-ray satellite transition energies of \ion{Fe}{25} compared with available calculations and experimental results. All energy values are expressed in eV, with line labels following the notation of~\cite{gabriel1969}. The CODATA2018~\citep{tiesinga2021} recommended value of $hc$ was used to convert literature values from \AA~to eV. Values in round parentheses below the theoretical values denote the absolute differences from the present FAC calculations.}
    \begin{tabular}{|c|l|cc|cccccc|}
    \hline
    Line  &  \multicolumn{1}{c|}{Transition}  & \multicolumn{2}{c|}{Resonance energy} & \multicolumn{6}{c|}{X-ray satellite energy}\\
\cline{3-10}          &       & This work$^a$ & B92$^b$ & This work$^a$ & Y12$^c$ & Y18$^d$ & S18$^e$ & VS$^f$ & Previous Experiments\\

    \hline$a$   &  $1s2p^2(^3\!P)\,^2\!P_{3/2} \to 1s^22p_{3/2}\,^2\!P^o_{3/2}$    & 4679.1 & 4677.0 & 6658.09 & 6658.12(1) & 6658.12(3) & 6659.37 & 6659.73 & 6658.41(107)$^g$\\
          &       &       & (2.10) &       & (-0.03) & (-0.03) & (-1.28) & (-1.64) & (-0.32) \\
    \hline$b$   &  $1s2p^2(^3\!P)\,^2\!P_{3/2} \to 1s^22p_{1/2}\,^2\!P^o_{1/2}$    & 4679.1 & 4677.0 & 6673.94 & 6674.08(1) & 6674.08(3) & 6675.15 & 6675.51 &  \\
          &       &       & (2.10) &       & (-0.15) & (-0.15) & (-1.21) & (-1.57) &  \\
    \hline$c$   &  $1s2p^2(^3\!P)\,^2\!P_{1/2} \to 1s^22p_{3/2}\,^2\!P^o_{3/2}$    & 4660.8 & 4658.6 & 6639.86 & 6639.86(1) & 6639.86(3) & 6640.82 & 6641.89 &  \\
          &       &       & (2.20) &       & (0.01) & (0.00) & (-0.96) & (-2.03) &  \\
    \hline$d$   &  $1s2p^2(^3\!P)\,^2\!P_{1/2} \to 1s^22p_{1/2}\,^2\!P^o_{1/2}$    & 4660.8 & 4658.6 & 6655.71 & 6655.82(1) & 6655.83(3) & 6656.87 & 6657.58 &  \\
          &       &       & (2.20) &       & (-0.11) & (-0.12) & (-1.16) & (-1.88) &  \\
    \hline$e$   &  $1s2p^2(^3\!P)\,^4\!P_{5/2} \to 1s^22p_{3/2}\,^2\!P^o_{3/2}$    & 4639.6 & 4639.0 & 6618.66 & 6620.35(1) & 6620.31(3) & 6621.67 & 6622.73 & 6621.46(124)$^g$ \\
          &       &       & (0.60) &       & (-1.69) & (-1.65) & (-3.01) & (-4.07) & (-2.80) \\
    \hline$f$   &  $1s2p^2(^3\!P)\,^4\!P_{3/2} \to 1s^22p_{3/2}\,^2\!P^o_{3/2}$    & 4633.5 & 4632.9 & 6612.52 & 6614.4(1) & 6614.36(3) & 6615.31 & 6616.73 &  \\
          &       &       & (0.60) &       & (-1.88) & (-1.84) & (-2.79) & (-4.20) &  \\
    \hline$g$   &  $1s2p^2(^3\!P)\,^4\!P_{3/2} \to 1s^22p_{1/2}\,^2\!P^o_{1/2}$    & 4633.5 & 4632.9 & 6628.37 & 6630.37(1) & 6630.33(3) & 6631.23 & 6632.30 &  \\
          &       &       & (0.60) &       & (-2.00) & (-1.96) & (-2.87) & (-3.93) &  \\
    \hline$h$   &  $1s2p^2(^3\!P)\,^4\!P_{1/2} \to 1s^22p_{3/2}\,^2\!P^o_{3/2}$    & 4625.3 & 4624.6 & 6604.28 & 6606.05(1) & 6606.01(3) & 6607.20 & 6608.61 &  \\
          &       &       & (0.70) &       & (-1.76) & (-1.73) & (-2.92) & (-4.33) &  \\
    \hline$i$   &  $1s2p^2(^3\!P)\,^4\!P_{1/2} \to 1s^22p_{1/2}\,^2\!P^o_{1/2}$    & 4625.3 & 4624.6 & 6620.13 & 6622.01(1) & 6621.98(3) & 6623.09 & 6624.15 &  \\
          &       &       & (0.70) &       & (-1.88) & (-1.85) & (-2.96) & (-4.02) &  \\
    \hline$j$   &  $1s2p^2(^1\!D)\,^2\!D_{5/2} \to 1s^22p_{3/2}\,^2\!P^o_{3/2}$    & 4665.3 & 4664.1 & 6644.34 & 6644.67(1) & 6644.64(3) & 6646.16 & 6646.52 & 6645.24(43)$^g$ \\
          &       &       & (1.20) &       & (-0.33) & (-0.30) & (-1.83) & (-2.18) & (-0.90) \\
    \hline$k$   &  $1s2p^2(^1\!D)\,^2\!D_{3/2} \to 1s^22p_{1/2}\,^2\!P^o_{1/2}$    & 4659.8 & 4658.1 & 6654.62 & 6654.79(1) & 6654.76(3) & 6656.15 & 6656.87 & 6654.19(71)$^g$ \\
          &       &       & (1.70) &       & (-0.16) & (-0.13) & (-1.53) & (-2.25) & (0.43) \\
    \hline$l$   &  $1s2p^2(^1\!D)\,^2\!D_{3/2} \to 1s^22p_{3/2}\,^2\!P^o_{3/2}$    & 4659.8 & 4658.1 & 6638.78 & 6638.82(1) & 6638.79(3) & 6640.47 & 6641.18 &  \\
          &       &       & (1.70) &       & (-0.04) & (-0.01) & (-1.69) & (-2.40) &  \\
    \hline$m$   &  $1s2p^2(^1\!S)\,^2\!S_{1/2} \to 1s^22p_{3/2}\,^2\!P^o_{3/2}$    & 4699.4 & 4697.7 & 6678.38 & 6677.65(2) & 6677.62(3) & 6679.82 & 6679.46 & 6676.84(72)$^g$ \\
          &       &       & (1.70) &       & (0.73) & (0.76) & (-1.44) & (-1.08) & (1.54) \\
    \hline$n$   &  $1s2p^2(^1\!S)\,^2\!S_{1/2} \to 1s^22p_{1/2}\,^2\!P^o_{1/2}$    & 4699.4 & 4697.7 & 6694.23 & 6693.62(2) & 6693.58(3) & 6695.70 & 6694.97 &  \\
          &       &       & (1.70) &       & (0.61) & (0.64) & (-1.47) & (-0.75) &  \\
    \hline$o$   &  $1s2s^2\,^2\!S_{1/2} \to 1s^22p_{3/2}\,^2\!P^o_{3/2}$           & 4555.6 & 4553.4 & 6534.67 & 6536.37(3) & 6536.3(3) & 6537.18 & 6538.22 & 6536.49(69)$^g$ \\
          &       &       & (2.20) &       & (-1.71) & (-1.64) & (-2.52) & (-3.55) & (-1.83) \\
    \hline$p$   &  $1s2s^2\,^2\!S_{1/2} \to 1s^22p_{1/2}\,^2\!P^o_{1/2}$           & 4555.6 & 4553.4 & 6550.51 & 6552.34(3) & 6552.27(3) & 6553.08 & 6553.77 & 6552.42(87)$^g$ \\
          &       &       & (2.20) &       & (-1.83) & (-1.76) & (-2.57) & (-3.26) & (-1.91) \\
    \hline$q$   &  $1s2s2p(^3\!P^o)\,^2\!P^o_{3/2} \to 1s^22s\,^2\!S_{1/2}$        & 4618.1 & 4615.3 & 6662.45 & 6662.19(1) & 6662.19(3) & 6663.31 & 6664.03 & 6662.09(54)$^g$, 6662.24(7)$^h$ \\
          &       &       & (2.80) &       & (0.26) & (0.26) & (-0.86) & (-1.58) & (0.36)$^g$, (0.21)$^h$ \\
    \hline$r$   &  $1s2s2p(^3\!P^o)\,^2\!P^o_{1/2} \to 1s^22s\,^2\!S_{1/2}$        & 4608.9 & 4604.9 & 6653.26 & 6652.78(2) & 6652.79(3) & 6653.65 & 6655.08 & 6654.19(71)$^g$, 6652.83(7)$^h$ \\
          &       &       & (4.00) &       & (0.49) & (0.47) & (-0.39) & (-1.82) & (-0.93)$^g$, (0.43)$^h$ \\
    \hline$s$   &  $1s2s2p(^1\!P^o)\,^2\!P^o_{3/2} \to 1s^22s\,^2\!S_{1/2}$        & 4635.8 & 4633.2 & 6680.10 & 6679.11(6) & 6679.1(3) & 6680.90 & 6681.26 &  \\
          &       &       & (2.60) &       & (0.99) & (1.00) & (-0.81) & (-1.17) &  \\
    \hline$t$   &  $1s2s2p(^1\!P^o)\,^2\!P^o_{1/2} \to 1s^22s\,^2\!S_{1/2}$        & 4632.6 & 4631.2 & 6676.94 & 6676.13(5) & 6676.11(3) & 6678.02 & 6678.38 & 6676.84(72)$^g$, 6676.20(7)$^h$ \\
          &       &       & (1.40) &       & (0.81) & (0.84) & (-1.08) & (-1.44) & (0.11)$^g$, (-0.74)$^h$ \\
    \hline$u$   &  $1s2s2p(^3\!P^o)\,^4\!P^o_{3/2} \to 1s^22s\,^2\!S_{1/2}$        & 4570.9 & 4570.1 & 6615.20 & 6616.56(1) & 6616.55(3) & 6617.78 & 6618.84 & 6617.89(124)$^g$, 6616.63(7)$^h,i$ \\
          &       &       & (0.80) &       & (-1.36) & (-1.35) & (-2.58) & (-3.64) & (-2.69)$^g$, (-1.43)$^h,i$ \\
    \hline$v$   &  $1s2s2p(^3\!P^o)\,^4\!P^o_{1/2} \to 1s^22s\,^2\!S_{1/2}$        & 4567.1 & 4566.3 & 6611.45 & 6612.85(1) & 6612.84(3) & 6613.90 & 6614.96 &  6612.94(7)$^i$ \\
          &       &       & (0.80) &       & (-1.40) & (-1.39) & (-2.45) & (-3.51) & (-1.49)$^i$ \\
    \hline
    \end{tabular}%
  \label{tab:theo_comp}%
{\begin{flushleft}
\footnotesize
\tablenotemark{$^a$}{This work}:~Present FAC calculation; {$^b$}{B92}:~\cite{beiersdorfer1992a}; {$^c$}{YS12}:~\cite{yerokhin2012}; {$^d$}{YS18}:~\cite{yerokhin2018}; {$^e$}{S18}:~\cite{sardar2018}; {$^f$}{VS}:~\cite{vainstein1978,beiersdorfer1992a}; {Previous experiments}:{$^g$}~\cite{beiersdorfer1993}, $^h$~\cite{rudolph2013}, and $^i$~\cite{steinbruegge2022}.\\
\end{flushleft}}
\end{table*}%

{Theoretical studies have extensively examined the KLL DR resonances of \ion{Fe}{25}, e.g.,~\cite{dubau1979,chen1986,nilsen1988,beiersdorfer1992a,nahar2001,gu2003dr,sardar2018,shah2018}. However, calculations for KL$n>2$ DR, and other lower charge states of Fe contributing to the K-shell X-ray lines remain scarce. In this work, we used FAC to compute all possible KL$n$ DR channels using a fully relativistic configuration interaction method, treating continuum states within the relativistic distorted wave approximation (see \cite{gu2008} and references therein). The independent-resonance approximation model is used to evaluate the DR resonance energies and strengths.}

For this, we included sets of initial configurations as [$1s^2\,(nl)^e$, $1s^1\,(nl)^{e+1}$, $1s^1\,(nl)^e\,n^\prime l^\prime$, and $1s^2\,(nl)^{e-1}\,n^\prime l^\prime$], sets of intermediate doubly-excited configurations as [$1s^1\,(nl)^{e+2}$ and $1s^1\,(nl)^{e}\,n^\prime l^\prime$], and sets of final configurations as [$1s^2\,(nl)^{e+1}$ and $1s^2\,(nl)^{e}\,n^\prime l^\prime$] in our structure calculations. The number of electrons, $e$, ranges from zero to four, covering \ion{Fe}{25} through \ion{Fe}{21} ions. Furthermore, $n$ has been limited to 2, since the maximum electron beam energy in the present experiment is below 7 keV. $n^\prime$ values are included up to 15, and $l\,l^\prime$ up to 8. Higher ($l>8$) contributions are expected to be negligible~\citep{gu2003dr}.

{Table~\ref{tab:theo_comp} presents a comparison of our present FAC energy calculations with previously reported results. The available literature focuses primarily on KLL DR resonance energies of \ion{Fe}{25} and their satellite transition energies in \ion{Fe}{24}, limiting the scope of our comparison. The most advanced configuration interaction (CI) energy level calculations for \ion{Fe}{24} are available in the literature~\citep{yerokhin2012,yerokhin2018,azarov2023}, with accuracies surpassing the available experimental results~\citep{beiersdorfer1993,rudolph2013,steinbruegge2022}. FAC, in comparison to these advanced calculations, deviate up to $\sim\pm2$~eV on an absolute scale. However, the relative differences are on the order of 0.5--1 eV, likely due to differences in the calculations of ground state energy. Nonetheless, our FAC predictions agree within $\sim$0.02\% with these state-of-the-art calculations.}

{Experimental values measured using the crystal spectrometer at Princeton Large Torus tokamak~\citep{beiersdorfer1993} show deviations of up to 2--2.5$\sigma$ with respect to our FAC results. We note that some of the experimental line accuracies reported in this work are affected by line blends. The high-resolution synchrotron experiment~\citep{rudolph2013}, measured four \ion{Fe}{24} transitions, which, with an exception of $u$, agree with our FAC predictions within 1 eV. The present FAC in comparisons with older calculations by~\cite{vainstein1978} show differences of up to 2--4 eV, while more recent multiconfiguration Dirac-Fock (MCDF) calculations~\citep{sardar2018} differ by up to 2 eV. The resonance energies reported in~\cite{beiersdorfer1992a} are consistently lower than our FAC predictions, with deviations ranging from 0.6--4 eV. Our previous work~\citep{shah2018} also reported KLL DR resonance energies for \ion{Fe}{25} through \ion{Fe}{19} ions, and found an average deviation of 1.3 eV from our previous FAC calculations.}

{Beyond KLL transitions, no additional data are available in the literature. Thus, we provide complete energy level calculations for \ion{Fe}{25} through \ion{Fe}{21}, up to $n^\prime=15$, as machine-readable files in the electronic version of this paper. An excerpt is shown in Table~\ref{tab:mrt}.}

\begin{table*}
  \centering
  \caption{FAC calculated DR data for \ion{Fe}{25} ion.}
    \begin{tabular}{ccccccccc}
    \hline\hline
$n$ shell$^{a}$ & $|d\rangle$  & $|d\rangle$ & $|f\rangle$ & $|f\rangle$  & Branching & $E_{\mathrm{beam}}$ & $E_{\mathrm{photon}}$ & $S_{\mathrm{DR}}$ strengths \\
      & index$^{b}$ &  configuration$^{c}$ &  index$^{d}$ & configuration$^{e}$ & ratio$^{f}$ & (eV)$^{g}$ & (eV)$^{h}$ & (cm$^2$ eV)$^{i}$ \\
    \hline
2     & 271   & 1s+1(1)1.2p-2(0)1 & 98    & 1s+2(0)0.2p-1(1)1 & 9.8998E-01 & 4.6253E+03 & 6.6201E+03 & 6.9947E-23 \\
2     & 271   & 1s+1(1)1.2p-2(0)1 & 99    & 1s+2(0)0.2p+1(3)3 & 6.4186E-03 & 4.6253E+03 & 6.6043E+03 & 4.5350E-25 \\
2     & 270   & 1s+1(1)1.2s+1(1)2.2p+1(3)3 & 97    & 1s+2(0)0.2s+1(1)1 & 9.9999E-01 & 4.6181E+03 & 6.6625E+03 & 4.7348E-24 \\
2     & 273   & 1s+1(1)1.2p-1(1)0.2p+1(3)3 & 98    & 1s+2(0)0.2p-1(1)1 & 2.9514E-03 & 4.6335E+03 & 6.6284E+03 & 2.7332E-24 \\
2     & 273   & 1s+1(1)1.2p-1(1)0.2p+1(3)3 & 99    & 1s+2(0)0.2p+1(3)3 & 9.5362E-01 & 4.6335E+03 & 6.6125E+03 & 8.8311E-22 \\
2     & 277   & 1s+1(1)1.2p-1(1)2.2p+1(3)1 & 98    & 1s+2(0)0.2p-1(1)1 & 7.6737E-01 & 4.6608E+03 & 6.6557E+03 & 1.2175E-21 \\
2     & 277   & 1s+1(1)1.2p-1(1)2.2p+1(3)1 & 99    & 1s+2(0)0.2p+1(3)3 & 2.3053E-01 & 4.6608E+03 & 6.6399E+03 & 3.6575E-22 \\
2     & 265   & 1s+1(1)1.2s+2(0)1 & 98    & 1s+2(0)0.2p-1(1)1 & 5.0987E-02 & 4.5556E+03 & 6.5505E+03 & 8.8268E-21 \\
2     & 265   & 1s+1(1)1.2s+2(0)1 & 99    & 1s+2(0)0.2p+1(3)3 & 5.2411E-02 & 4.5556E+03 & 6.5347E+03 & 9.0733E-21 \\
2     & 278   & 1s+1(1)1.2p+2(4)5 & 99    & 1s+2(0)0.2p+1(3)3 & 5.9891E-01 & 4.6653E+03 & 6.6443E+03 & 2.7967E-19 \\
2     & 267   & 1s+1(1)1.2s+1(1)2.2p-1(1)3 & 97    & 1s+2(0)0.2s+1(1)1 & 9.8286E-01 & 4.5709E+03 & 6.6152E+03 & 5.7210E-22 \\
2     & 274   & 1s+1(1)1.2s+1(1)0.2p+1(3)3 & 97    & 1s+2(0)0.2s+1(1)1 & 7.1247E-03 & 4.6358E+03 & 6.6801E+03 & 1.7592E-21 \\
\hline\hline
\end{tabular}%
{\begin{flushleft}
\footnotesize
\tablenotemark{}{This table is published in its entirety in the electronic edition of the Astrophysical Journal. A portion is shown here for guidance regarding its form and content.}\\
\tablenotemark{$^a$}{Principal quantum number of the shell in which the free electron is recombining during the dielectronic capture process.}\\
\tablenotemark{$^b$}{Index for the intermediate doubly excited states $|d\rangle$.}\\
\tablenotemark{$^c$}{Relativistic shell configuration of the $|d\rangle$ states. Here, \texttt{2p+2(2)} represents two electrons in the $2p_{3/2}$ subshell with a total angular momentum $J = 1$, while \texttt{2p-2(2)} represents two electrons in the $2p_{1/2}$ subshell with $J = 1$. The number in parentheses indicates two times the total angular momentum of the coupled shell. Following the parentheses, a number representing the $2J$ value when all preceding shells are coupled is included. Therefore, \texttt{2p+2(2)2.2p-2(2)0} represents a state described as $[(2p_{3/2})^2_{(J = 1)} (2p_{1/2})^2_{(J = 1)}]_{J = 0}$.}\\
\tablenotemark{$^d$}{Index for the final excited states $|f\rangle$. For each $|d\rangle$ state, there are several $|f\rangle$ states.}\\
\tablenotemark{$^e$}{Relativistic shell configuration of the $|f\rangle$ states.}\\
\tablenotemark{$^f$}{DR branching ratio.}\\
\tablenotemark{$^g$}{DR resonance energy in electron volts. This is also the electron beam energy required to resonantly excite the $|d\rangle$ state.}\\
\tablenotemark{$^h$}{DR satellite photon energy in electron volts.}\\
\tablenotemark{$^i$}{DR resonance strength in cm$^2$ eV.}\\
\end{flushleft}
}
\label{tab:mrt}%
\end{table*}%

\begin{figure*}
    \centering
    \includegraphics[width=0.95\linewidth]{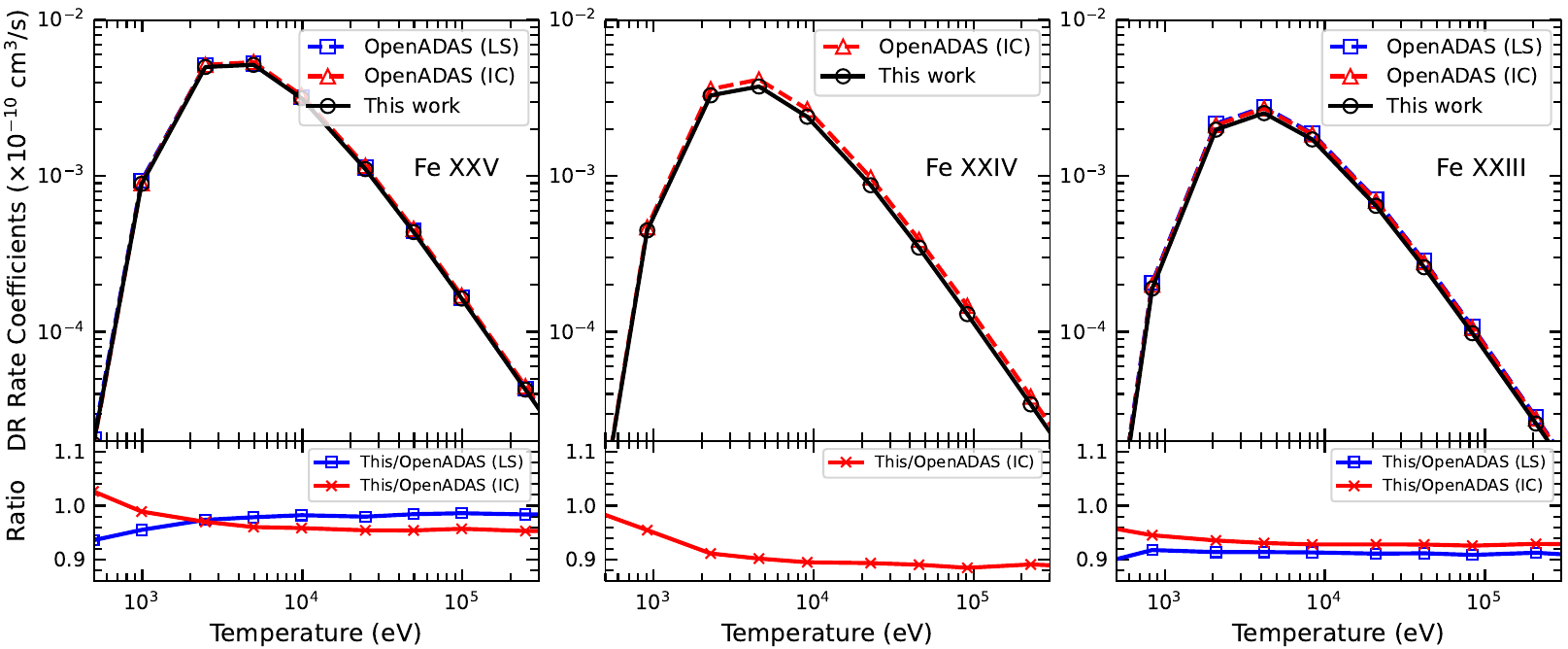}
    \caption{{K-shell DR rate coefficients of \ion{Fe}{25}-\ion{}{23} ions as a function of plasma electron temperature. The top panels show the comparison between the present DR rates and those available in the OPEN-ADAS database in their LS and IC coupling formats. The bottom panels display the ratio between the present and OPEN-ADAS rate coefficients.}}
    \label{fig:rates}
\end{figure*}

{To determine DR resonance strengths, we also used FAC to calculate dielectronic capture strengths, radiative and autoionization transition rates, and branching ratios. All these values for ions under this study are provided in the machine-readable files. Table~\ref{tab:mrt} lists partial KL$n$ DR resonance data for \ion{Fe}{25} ion, including their intermediate doubly excited and final state configurations, as well as their resonance energies, photon energy, branching ratio, and strengths. These data could potentially support comparisons and updates to atomic data incorporated into spectral models such as AtomDB~\citep{foster2012}, SPEX~\citep{kaastra1996}, and CHIANTI~\citep{delzanna2015}.}

{Our theoretical data also facilitate the calculation the K-shell DR rate coefficient, which is an essential quantity for modeling collisionally-ionized plasmas found in astrophysical and laboratory fusion plasma environments. Using the rate coefficient formula from~\cite{gu2003dr}, we compute DR rates and compare them with data from the widely used OPEN-ADAS database~\footnote{\url{https://open.adas.ac.uk/}}. The OPEN-ADAS rates, mainly derived from AUTOSTRUCTURE~\citep{badnell1986}, are available in both Russell-Saunders LS coupling and intermediate IC coupling formats for \ion{Fe}{25}\citep{bautista2007}, \ion{Fe}{24}\citep{colgan2004}, and \ion{Fe}{23}\citep{colgan2003}. We note that these rates are also used in the APEC/AtomDB~\citep{foster2012} spectral model used in the X-ray astrophysics community. Figure~\ref{fig:rates} compares our derived DR rate coefficients with those in the OPEN-ADAS database for electron temperatures ranging from 0.5 to 200 keV. Our calculated rates for \ion{Fe}{24} and \ion{Fe}{23} are slightly lower compared to those from OPEN-ADAS, whereas we find good agreement for \ion{Fe}{25}. Overall, FAC and OPEN-ADAS DR rates agree within 10\%.}

{For comparison with our present EBIT experiment, as shown in our previous work~\citep{shah2019}, we fed atomic data into the collisional-radiative model of FAC, which solves  population balance equations on a fine-grid of the beam energy, matching the present experimental conditions~\citep{gu2008}. The resulting level populations enter the calculations of total line emission cross sections from both DR and EIE processes as a function of electron beam energy. Additionally, since we observed X-rays following radiative recombination (RR) into the L and M shells of Fe ions, we calculate total RR cross sections with FAC, finding agreement within $\sim$4\% with those previously reported by~\cite{chen2005}.}

{The unidirectional electron beam in an EBIT leads to polarized and anisotropic X-ray emission following RR and DR processes~\citep{oppenheimer1927,henderson1990,beiersdorfer1996,chen2005,shah2015}. Since our SDD detector is mounted at 90$^{\circ}$ with respect to the electron beam propagation axis, observed RR and DR X-ray intensities require corrections for polarization effects. Using FAC, we calculate polarization correction factors, accounting for depolarization effects arising from the cyclotron motion of electrons within the electron beam and radiative cascades~\citep{gu1999}. For the former, we apply a transverse energy component of $\sim$450~eV, measured in our previous study~\citep{shah2018}, to our FAC polarization calculations.}

\section{Data Analysis and Discussion of the Results}\label{result}

\subsection{K-shell cross sections: KL$n$ DR satellites}\label{sec:cross}

In addition to determining the DR line energies, we also infer the cross sections for all the observed satellite lines and compare them with predictions. To do this, we need to convert the observed X-ray counts into absolute cross sections. However, due to the unknown electron beam density, ion-number density, and overlap factor between electron beam and ion cloud, absolute cross sections cannot be directly determined. Nonetheless, as shown by~\citet{knapp1989} and~\citet{beiersdorfer1992a}, we can use the theoretical cross sections of RR into $n=2$, observed in this experiment, to normalize the DR satellites. 

For this procedure, the charge-state distribution must be known. However, strong DR resonances modify it. In principle, this distribution can be estimated by comparing the experimental data with theoretical DR resonance strengths for each charge state, assuming a steady-state charge balance is achieved during a sufficiently slow electron beam-energy scan. It is known that when the scan frequency is faster than the recombination and ionization rates, the charge balance remains stable, resulting in minimal oscillations in the ion population. Conversely, when it is lower than those rates, oscillations in the ion population become large.

\begin{figure}
    \centering
    \includegraphics[width=\columnwidth]{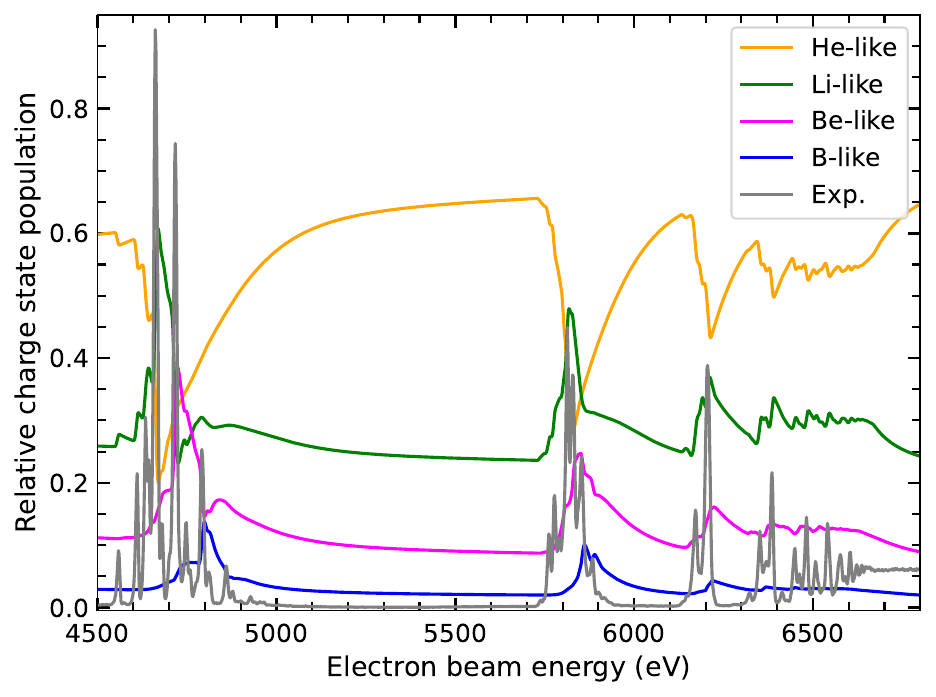}
    \caption{Predicted charge-state distribution of Fe ions trapped under the present experimental conditions. {Summed DR satellite intensities within the K$\alpha$ cut are superimposed to highlight the depletion of certain charge states caused by strong DR resonances at different resonance energies.}}
    \label{fig:dither}
\end{figure}

{In the present experiment, a slow scan frequency of 50~eV s$^{-1}$ was employed due to technical limitations of the drift tube high voltage power supply}. This means that when the electron beam passes through the strongest helium-like KLL resonance, which has a width of $\sim$7~eV, it spends about 140~ms on it. The strongest peak DR cross section is $\sim5\times10^{-20}$ cm$^2$. During this 140~ms period, the helium-like ion population will approximately decay according to $\exp(-n_e \, \sigma_e \, v_e \, dt)$, where $n_e$ is the electron beam density, $\sigma_e$ is the cross section, $v_e$ is the electron beam velocity, and $dt$ is the DR peak scan time. Assuming an effective electron density---i.e., considering the ion-electron beam fractional overlap factor---of approximately $10^{10}$ cm$^{-3}$, the helium-like population decreases by a factor of $\exp(-0.28)$, after scanning through the strongest DR resonance. Subsequently, the helium-like population recovers by electron-impact ionization at slower rate. This complicates the comparison between experimental and theoretical resonance strengths as a function of the beam energy. To avoid this, we must determine the charge-state distribution in the present experiment.

Following the approach outlined by~\citet{penetrante1991}, we simulate the charge-state distribution by numerically solving 27 coupled differential equations, each of them describing the abundance of each charge state, with their ionization and recombination cross sections calculated by FAC. The simulation takes into account experimental parameters such as electron-beam energy and current, neutral gas density, and applied axial trap potential. We tested such simulations against experiments in our previous works~\citep{grilo2021,beila2023,grilo2024}. The simulated charge-state distribution during the upward scan is shown in Fig.~\ref{fig:dither}. {As expected, it shows an overall increase of the helium-like abundance with growing electron beam energy, and an instantaneous drop on the DR resonance with a corresponding jump of the lithium-like population. Similar patterns are observed for the other charge states studied in the present experiment.}

\begin{figure*}
    \centering
    \includegraphics[width=\linewidth]{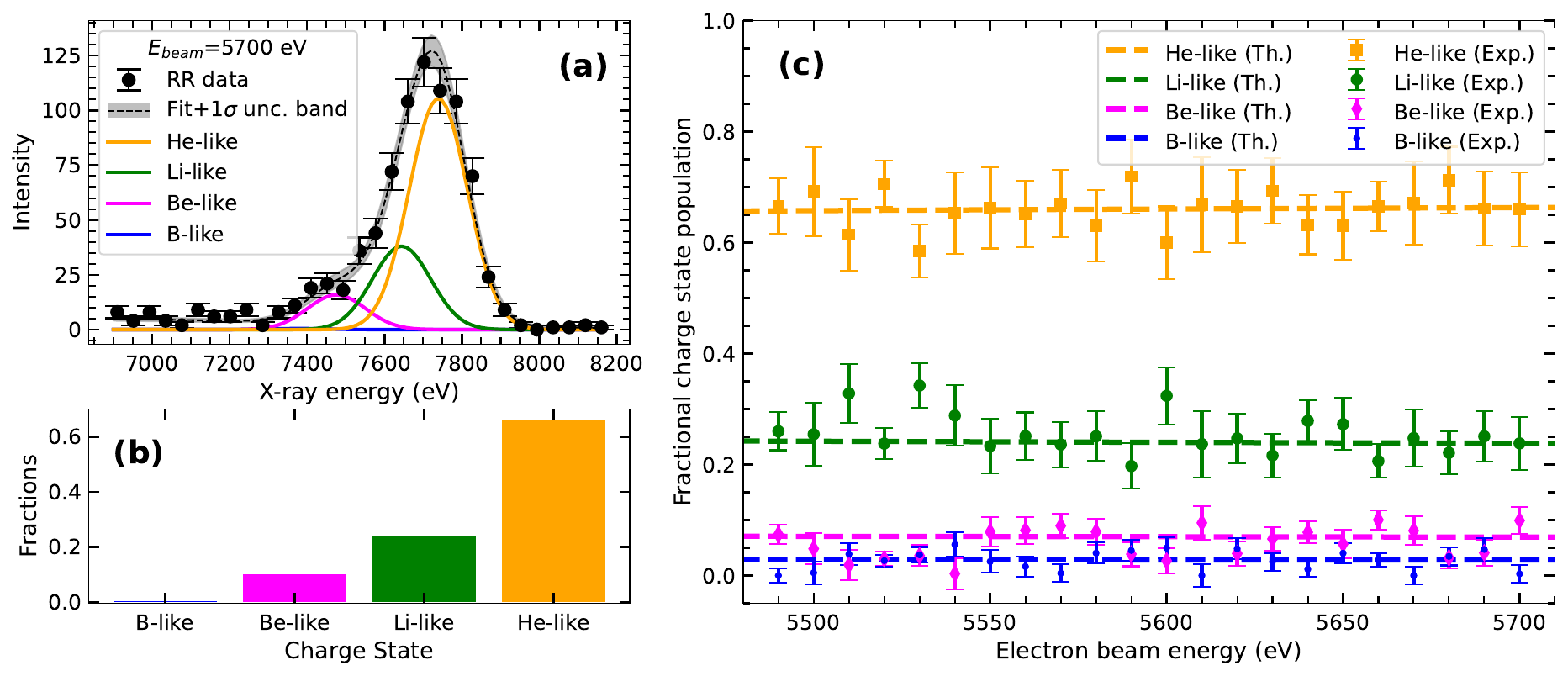}
    \caption{(a) Example fit to the radiative recombination (RR) into $n=2$ data obtained from a 10-eV broad ROI around an electron-beam energy of 5700 eV. (b) Fractional populations of charge states derived from this fit. (c) Derived fractional charge-state populations as a function of electron-beam energy based on the example in (a) and (b). The extracted ion populations are compared with the theoretical charge balance simulation shown in Fig.~\ref{fig:dither}.}
    \label{fig:expcs}
\end{figure*}

\begin{figure*}
    \centering
    \includegraphics[width=1.15\linewidth,angle=90]{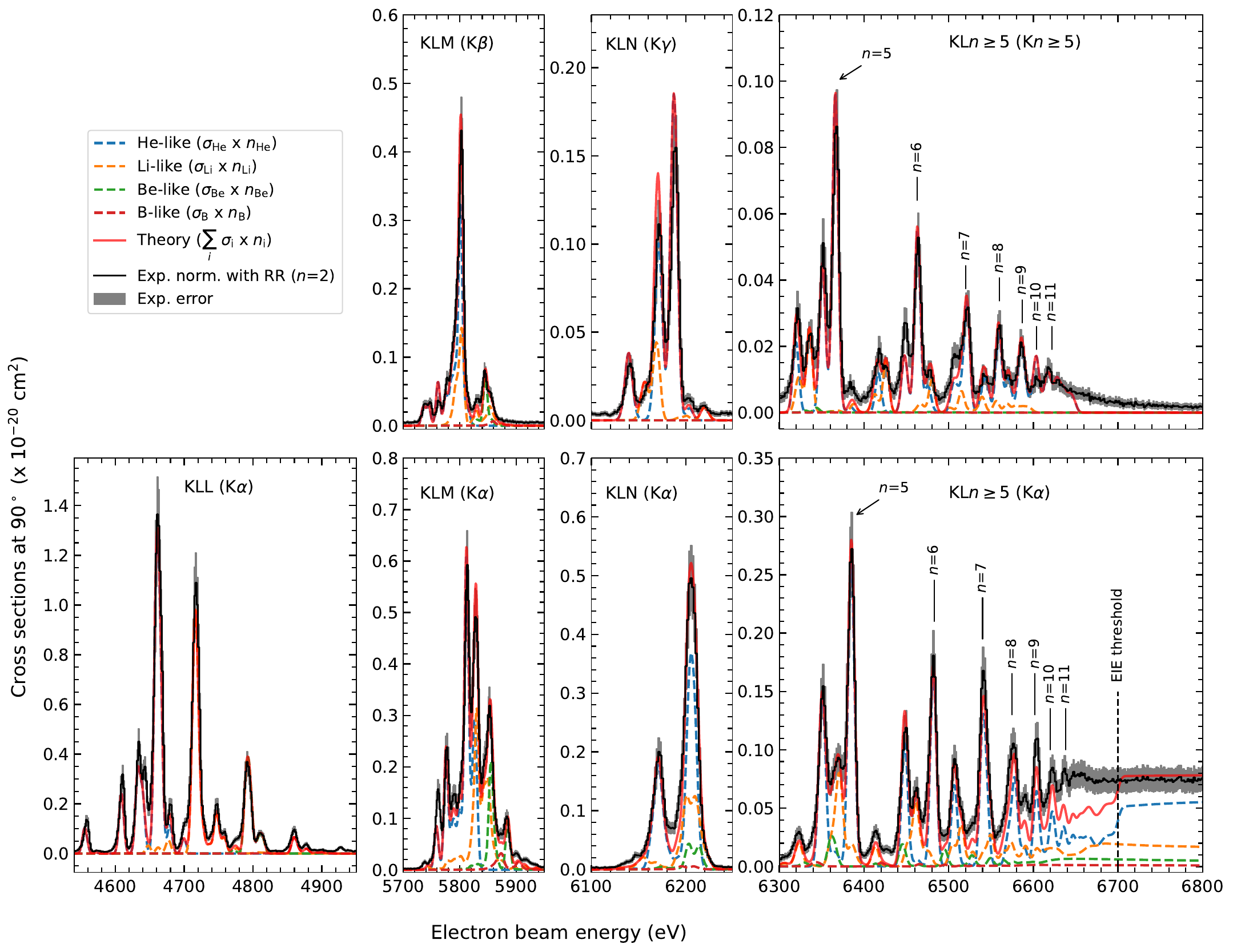}
    \caption{Total DR cross sections (black curve) measured at an angle of 90$^\circ$ with respect to the electron beam propagation axis and normalized to RR ($n=2$) versus electron-beam energy, with the total uncertainty shown as gray band. Distorted wave predictions obtained with FAC for present charge-state distribution (see Fig.~\ref{fig:dither}) is shown as red curve. Different charge state components are also shown as dashed lines. Top three panels: Projection of the KL$n$ DR satellites from the RR ($n=2$) slice in Fig.~\ref{fig:exp2d}. Bottom four panels: Projection of the K$\alpha$ slice from Fig.~\ref{fig:exp2d}.}
    \label{fig:cross90}
\end{figure*}

{The charge-state distribution does not change substantially in regions where no DR resonances are present, such as below the KLL, between the KLL and KLM resonances, and between KLM and KLN (see Fig.~\ref{fig:exp2d}). We first correct the intensities in 2D plot for the cyclically varying electron beam current density. Then we extract projections of the spectrum along the RR ($n=2$) band in these regions for cross-section calibration. Specifically, we first take 10-eV slices of data between KLL and KLM in the electron-beam energy range of 5480--5720 eV and project the summed spectrum onto the X-ray axis, as demonstrated in Fig.~\ref{fig:expcs}(a). The observed RR intensity depends on both the RR cross sections at 90$^\circ$ with respect to the electron beam propagation axis and the ion abundances in the EBIT. By calculating the angle-dependent RR cross sections, we can estimate the fractional charge-state distribution in the trap from this data, as shown in previous works~\citep{knapp1989,beiersdorfer1992a,smith2000,chen2005,shahharman2019,shah2021}.}

{The RR X-ray energies are the sum of the electron-beam energy and the binding energy of the ion state into which the electron recombines. For each slice of the electron beam energy, we first calculate the RR X-ray energies and 90$^\circ$ RR cross sections using FAC. These calculations include the fine-structure component of each ion. The effective electron-beam energy differs from nominal because of the negative space charge of the electron beam compensated by the positive space charge of the ion cloud~\citep{currell2005}. We therefore calibrated the electron beam energy using theoretical DR resonance energies. Using our calculated RR cross sections, we generate synthetic RR spectra for each ion by convolving the FAC-RR data on the photon-energy axis with a Gaussian function with 120-eV FWHM to account for the resolution of the SDD. The centroids and amplitudes of the synthetic spectrum for each charge state are then used as initial parameters for fitting the experimental RR data. In the fit, we fix the relative difference between peak centroids and share the widths of each peak, while allowing the amplitudes to be free parameters. This allows us to obtain the relative charge balance of the trapped ions by taking the ratios of the fitted RR intensities and cross sections for each charge state. The resulting fit and the derived fractional charge-state distribution are shown for one energy slice in Fig.~\ref{fig:expcs}(a) and (b). We repeat this procedure for all RR data slices to extract the charge-state distribution as a function of electron-beam energy. The results are in good agreement with our charge balance simulations, as shown in Fig.~\ref{fig:expcs}(c), which further strengthens our confidence in these simulations.}

{Given that RR cross sections can be predicted within 3--5\%~\citep{knapp1989, chen2005}, and the charge-state distribution is experimentally known, we can now derive a normalization factor for each charge state of Fe and infer DR cross sections, as demonstrated in~\cite{shahharman2019,shah2021}. However, predictions suggest that even with high electron-beam energy resolution of $\sim$7 eV, blending between resonances of different charge states persist. Thus, normalizing each charge state individually is challenging. Instead, we use the total RR counts divided by the product of fractional abundances and RR cross sections, i.e., an effective normalization factor defined as 
$I_\mathrm{RR}^{\mathrm{total}}/(\sum_{cs} n_{cs} \times \sigma^\mathrm{RR}_{cs})$,
where ($cs$) indicates the charge state. Each RR slice in the KLL-KLM region gives the effective normalization factor. All factors are consistent within their statistical error bars. Thus, we took the weighted average of the normalization factor for the KLL-KLM range, which was found to be $(3.47 \pm 0.28) \times 10^{22}~\mathrm{counts~cm}^{-2}$.}

{The RR fit procedure is sensitive to uncertainties in the overall energy scale of the RR peaks, which are affected both by the energy scale calibration of the SDD as well as by the effective space charge of the electron beam. Both of these can be modeled as a shift in the global energy scale of the RR complex, retaining the relative peak energies predicted by theory. We conservatively take the uncertainty on this energy scale as 10 eV, and derive a systematic uncertainty of $\sim$4.5\% on the normalization factor by shifting the energy scale and refitting the peaks. Furthermore, we considered a 5\% uncertainty in the theoretical RR cross sections used for the normalization procedure. We add statistical and systematic uncertainties in quadrature and derive the effective normalization factor to be $(3.47 \pm 0.37) \times 10^{22}~\mathrm{counts~cm}^{-2}$. This value is then used to normalize measured DR intensities as a function of electron beam energy, shown in Fig.~\ref{fig:cross90}.}

{In addition to the RR $n=2$ band between KLL and KLM, we apply the above-mentioned procedure for the RR band below KLL in the electron beam energy range of 4300--4500 eV, obtaining a normalization factor of $(3.48 \pm 0.34) \times 10^{22}~\mathrm{counts~cm}^{-2}$. The normalization factor for the RR band between KLM and KLN in the beam energy range of 5900--6100 eV is $(3.6 \pm 0.40) \times 10^{22}~\mathrm{counts~cm}^{-2}$. All normalization factors from different electron beam energy regions, ranging from 4300 eV to 6100 eV, are consistent within their error bars, suggesting that the measurements are not significantly affected by variations in effective electron density and beam-ion overlap across the beam energy scan range and due to different charge states.}

Finally, we compared our measured DR cross sections with FAC predictions weighted by the simulated charge-state distribution as a function of electron-beam energy (see Fig.~\ref{fig:cross90}). 
{We observe overall good agreement for the set of DR satellites contributing to the collisionally excited K$\alpha$, K$\beta$, K$\gamma$, and K$n\geq5$ X-ray lines.
However, some discrepancies remain, such as the high-$n$ satellites near the excitation threshold for Fe K-shell lines, which are slightly underestimated by the FAC calculations. This may be partly due to the exclusion of resonances with $n^\prime > 15$ in our calculations. In addition, there are some differences in specific resonances: the KLN (K$\gamma$) resonances at $\sim$6170 and $\sim$6190 eV, where the FAC predictions slightly exceed the measurements, and the KLM (K$\alpha$) resonance at $\sim$5760 eV, where the FAC underestimates the cross sections. The reasons for these discrepancies are not yet clear. For beam energies above the excitation threshold ($\geq 6700$ eV), the measured electron impact cross sections for the Fe K$\alpha$ line show good agreement with the FAC predictions. We also observe pileup counts above the Fe K$\alpha$ satellites due to the high intensities of Fe L-shell photons near 1 keV X-ray energy (see above KLL resonances in Fig.~\ref{fig:exp2d}). This may contribute to the count enhancement in the K$\beta$ resonances as well as in the RR $n=2$ band. However, we find that this contribution is less than 0.5\% of the total counts, which means that it does not significantly affect our cross-section measurements.}

\begin{figure}
    \centering
    \includegraphics[width=\columnwidth]{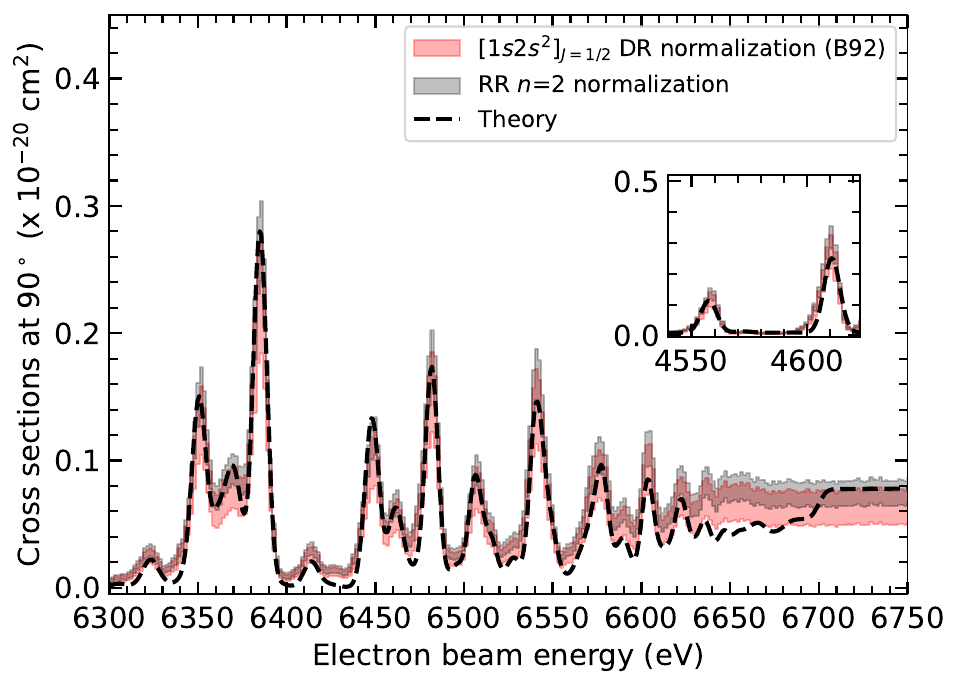}
    \caption{Comparison of present measurements normalized using one of the KLL resonances with the doubly excited state configuration, $[1s\,2s^2]_{J=1/2}$, as reported by~\citet{beiersdorfer1992a} (B92), with normalization using RR into $n=2$ as observed here. The inset shows the region of normalization line, $[1s\,2s^2]_{J=1/2}$ at $\sim$4554 eV (left peak). The dashed curve represents FAC predictions.}
    \label{fig:comp}
\end{figure}

We further validate our results by comparing them to the KLL DR satellite cross sections of \ion{Fe}{25} measured by \citet{beiersdorfer1992a} using the Lawrence Livermore National Laboratory EBIT (LLNL-EBIT) with an estimated uncertainty of 20\%, mostly resulting from the response function and  polarization sensitivity of the crystal spectrometer used there. For the comparison, we take the measured cross sections of the X-ray lines $o$ and $p$ underlying the first KLL DR resonance at an electron-beam energy of $\sim$4554 eV. This DR resonance is interesting because its doubly excited state, $[1s\,2s^2]{J=1/2}$ decays by two-electron-one-photon (TEOP) electric-dipole transitions with final states $[1s^2\,2p_{1/2,3/2}]_{J=1/2,3/2}$ for the $o$ and $p$ lines~\citep{zhang2004}, respectively.

Although this resonance is weak, it is well separated from the rest, and its doubly excited state with $J=1/2$ leads to unpolarized, isotropic X-ray emission~\citep{shah2018}, and only causes a small depletion of the helium-like ion abundance.
{We take the sum of the $o$ and $p$ cross sections, corrected for the spectrometer response factor $G(s/w)$ as given in \citet{beiersdorfer1992a}, and obtain a normalization factor of $(4.08\pm0.82)\times10^{22}~\mathrm{counts~cm^{-2}}$, with an uncertainty of approximately 20\% mostly arising from that of the LLNL-EBIT measurements~\citet{beiersdorfer1992a}. This value is also in good agreement with the value inferred above from the RR $n=2$ KLL-KLM energy band. The cross sections obtained from two independent normalization methods are compared in Fig.~\ref{fig:comp}, and are consistent within their uncertainty limits and in agreement with the distorted-wave predictions for DR and EIE. This further confirms the uncertainty estimates of our resulting cross sections, and comprehensively benchmarks FAC calculations involving RR, DR, and EIE processes.}

\subsection{L-shell cross sections: cascades following KL$n\geq3$ DR satellites}\label{sec:lcross}

\begin{figure*}
    \centering
    \includegraphics[width=0.95\linewidth]{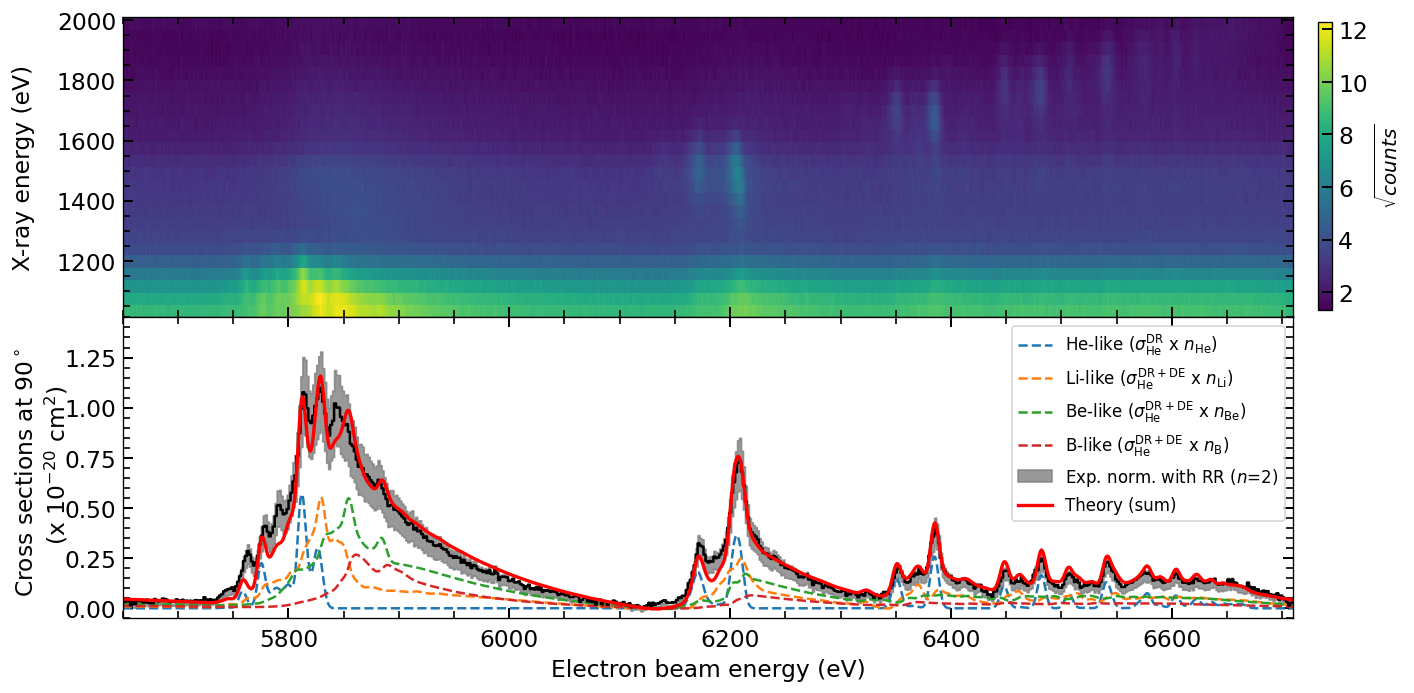}
    \caption{Top panel: 2D X-ray intensity histogram as a function of electron beam and X-ray energies. Bottom panel: Summed X-ray intensity projected onto the electron beam energy axis, normalized to the theoretical RR ($n=2$) cross sections and corrected for the transmission of the filter on the X-ray detector. The different line emission components for each charge state, as predicted by the FAC, are also shown.}
    \label{fig:l-shell}
\end{figure*}

Energetically below the primary K-shell DR satellite lines we detected low-energy radiative cascades from resonances {(e.~g.~$n=3\rightarrow2$)} where the spectator electron occupies the $n\geq3$ shell. In the case of lithium-like KLN resonances, K-series decays such as $n=4\rightarrow1$ and $n=2\rightarrow1$ appear as a diagonal series of bright spots (see Fig.~\ref{fig:exp2d}) from high-$n$ DR satellites alongside transitions such as $n=4\rightarrow2$, $n=4\rightarrow3$, and $n=3\rightarrow2$ at energies below $\sim$2~keV, as shown in Fig.~\ref{fig:l-shell}. As an example, the KLN resonances at $\sim$1.5~keV and $\sim$1.1~keV correspond to $n=4\rightarrow2$ and $n=3\rightarrow2$ decay paths, respectively. To obtain the cross sections for these low-energy cascades, we used our RR ($n=2$) normalization factor. We first corrected the observed intensity for the 1~$\mu$m-thick carbon foil in front of the X-ray detector using known transmission coefficients from~\citet{henke1993}. This is more important for the low-energy L-shell X-rays than for those above 6~keV from K-shell vacancies. In our previous study~\citep{shah2019}, we validated the filter transmission below 2 keV using K-shell EIE lines and RR lines of \ion{O}{8} and \ion{Ne}{10} ions, and found those coefficients agreeing within $\sim$3\%. We take into account this uncertainty in our normalization factor.

The continuous X-ray background seen around 1\,keV is primarily due to $n=3-2$ and $n=4-2$ direct EIE of \ion{Fe}{24}\,--\ion{}{20} ions. These EIE components are typically linear at high electron-beam energies. Here, however, they modulate due to the changes in charge-state distribution at strong K-shell DR resonances are seen (see~Fig.~\ref{fig:dither}). For instance, on the KLN resonances of helium-like, the abundance of lithium-like ions increases, and accordingly the $n=4\rightarrow2$ and $n=3\rightarrow2$ emissions from direct EIE of lithium-like ions. Outside the resonance, they are gradually ionized back to helium-like charge state, and the lithium-like EIE emissions diminish, as apparent at beam energies above the resonances in Fig.~\ref{fig:l-shell}.

{To model the L-shell spectrum, in addition to the DR cascades discussed in Sec.~\ref{sec:th}, we also computed with FAC direct EIE cross sections at 90 degrees for $n=3-2$ and $n=4-2$ transitions in \ion{Fe}{24}\,--\ion{}{20} ions as a function of electron beam energy. We compared our EIE predictions with previous experimental and theoretical results of~\cite{gu2001,chen2005} and found good agreement. The summed EIE and DR cross sections in the 1--2 keV range of X-ray energy, and weighted by the charge-state distribution of each ion, are shown in Fig.~\ref{fig:l-shell}. These low-energy satellite cascades from K-shell DR satellites are critical for the complete spectral plasma modeling, as blend with L-shell EIE lines for \ion{Fe}{25}\,--\ion{}{20} ions, affecting the Fe L-shell spectrum.}

\section{Summary and Conclusions}\label{concl}

We present the most comprehensive measurements to date of K-shell DR resonances for \ion{Fe}{25}\,--\ion{}{21} ions, resolving KL$n$ satellites up to $n^\prime$ = 11 with excellent electron-ion collision energy resolution and high counting statistics. This allowed us to determined cross sections for all observed DR satellites, normalized to those of RR into $n=2$, with total uncertainties in cross-section well below 15\%. We also compared these results with those from an~\textit{independent} normalization based on previous LLNL-EBIT measurements~\citep{beiersdorfer1992a}. We use the experimental cross sections to benchmark relativistic distorted-wave calculations performed using FAC~\citep{gu2008}, and found overall an excellent agreement. Our investigation of low-energy lines below 2 keV due to cascades from $n\geq3$ K-shell DR resonances and direct electron-impact excitation process also found good agreement with the total L-shell cross sections predicted with FAC. This work thus comprehensively benchmarks K-shell and L-shell FAC calculations for astrophysically abundant Fe ions and underscores the suitability of this code for efficiently generating the large atomic data sets required for accurate modeling of astrophysical and fusion plasmas.

Our experimental and theoretical data can immediately be used to interpret existing spectra observed by Hitomi, Chandra, XMM-Newton, and Suzaku, and will improve the analysis of high-resolution X-ray spectra currently from the X-ray microcalorimeter Resolve onboard the XRISM observatory~\citep{xrism2024cygx3,xrism2024ngc4151}, and in future from the planned Athena~\citep{pajot2018athena} mission. With our results, a rigorous check of DR data used in the leading spectral modeling codes SPEX~\citep{kaastra1996}, CHIANTI~\citep{delzanna2015}, and AtomDB~\citep{foster2012} could perhaps explain the discrepancies those codes shown in comparison with Hitomi observations~\cite{hitomiatomic2018}. In future, experiments with miniature EBITs~\citep{micke2018} capable of achieving twice as high electron beam energy resolution than the work presented here in combination with a high-resolution, wide-band X-ray microcalorimeter~\citep{porter2008,porter2009,smith2019,szypryt2019,gottardi2024} capable of covering both K- and L-shell bands with an exceptional 2--5 eV resolution will further deepen our understanding of the DR process as dominant line-formation mechanism in plasmas, and test atomic theory and plasma models more stringently than now possible. This is a crucial task towards fully exploiting the unique capabilities of present and upcoming X-ray observatories for understanding the role high energy X-ray sources in the evolution of galaxies, their clusters, and the intergalactic medium, where Fe emissions reveal the hottest components of the plasmas pervading them.

\section*{acknowledgements}
Research was funded by the Max-Planck-Gesselschaft (MPG), Germany. C.S. acknowledges support from MPG and NASA-JHU Cooperative Agreement. M.F.G. acknowledges support by NASA under APRA grant 80NSSC 20K0835. M.A.L. and F.S.P. acknowledge support from the NASA Astrophysics Program. P.A. and F.G. acknowledge support from Funda\c{c}\~{a}o para a Ci\^{e}ncia e Tec\-no\-lo\-gia, Portugal, under contracts No .~LA/P/0117/2020 (REAL) and UI/BD/151000/2021. This research has been carried out under the High-Performance Computing Chair—a R\&D infrastructure (based at the University of Évora; PI: M Avillez), endorsed by Hewlett Packard Enterprise (HPE), and involving a consortium of higher education institutions (University of Algarve, University of Évora, NOVA University of Lisbon, and University of Porto), research centres (CIAC, CIDEHUS, CHRC), enterprises (HPE, ANIET, ASSIMAGRA, Cluster Portugal Mineral Resources, DECSIS, FastCompChem, GeoSense, GEOtek, Health Tech, Starkdata), and public/private organizations (Alentejo Tourism-ERT, KIPT Colab). 
%

\bibliographystyle{aasjournal}
\bibliography{references}
\end{document}